\documentclass[twocolumn,pra,aps,showpacs]{revtex4-1}

\usepackage{amsmath,amssymb,amsfonts,bbm,graphicx}
\usepackage{amstext}
\usepackage{color}

\newcommand{\ket}[1]{| #1 \rangle}
\newcommand{\bra}[1]{\langle #1| }

\newcommand{\LM}{_\textit{LM}}
\newcommand{\MR}{_\textit{MR}}
\newcommand{\co}{^\textrm{(co)}}

\begin{document}

\title{Spatial adiabatic passage via interaction-induced band separation}

\author{Albert Benseny}
\email{albert.benseny-cases@oist.jp}
\author{J\'er\'emie Gillet}
\author{Thomas Busch}
\affiliation{Quantum Systems Unit, OIST Graduate University, Onna, Okinawa 904-0495, Japan}

\date{\today}

\begin{abstract}
The development of advanced quantum technologies and the quest for a deeper understanding of many-particle quantum mechanics requires control over the quantum state of interacting particles to a high degree of fidelity.
However, the quickly increasing density of the spectrum, together with the appearance of crossings in time-dependent processes, makes any effort to control the system hard and resource intensive.
Here we show that in trapped systems regimes can exist, in which isolated energy bands appear that allow to easily generalize known single-particle techniques.
We demonstrate this for the well-known spatial adiabatic passage effect, which can control the center-of-mass state of atoms with high fidelity.
 \end{abstract}

\pacs{03.75.Lm, 03.75.Be, 42.50.Dv}

\maketitle

\section{Introduction}

Understanding the effects of interactions between many particles at the quantum level is an important task for increasing the access to, and control over, ever larger parts of the Hilbert space \cite{Jaksch:99,Sorensen:01}. However, this is a difficult problem, as interacting many-particle states are usually too complex to allow for exact analytical treatment and often require numerical resources that are out of reach for current computers.
One way to approach this problem is to study small systems first and use the developed understanding for scaling up.
This allows theoretical treatment and experimental verification, as recent  progress in cold atomic gases has led to control over the trapping of small numbers of particles and the ability to measure them with high precision \cite{Murmann:15}.
Using Feshbach resonances one can then change the interaction strength between the particles and engineer well-defined quantum states. The importance of this lies in the ability to explore new aspects of fundamental quantum mechanics as well as to design new applications in quantum technologies. 

However, no generic and straightforward strategies exist to develop new engineering tools that take advantage of the full Hilbert space.
One controlled way to make progress is to try to generalize known techniques for single-particle states to either weakly-correlated many-body states or to small strongly-correlated systems~\cite{cao:11}.
In this work we focus on the well-known spatial adiabatic passage (SAP) protocol~\cite{eck:04,gre:04,Longhi:06,MenchonSound:14}, which is a technique that allows to transfer a localized wave function between different positions in space by adiabatically following a specific energy eigenfunction.
SAP for ultracold atoms has up to now only been studied for a single atom~\cite{eck:04,Loiko:11,Morgan:13,men:14},
weakly-interacting systems~\cite{Graefe:06,Rab:08,ben12,bradly12},
or fermionized bosons~\cite{Benseny:10}.
Here we discuss a system of two interacting bosons in a triple well potential and, using exact solutions and Hubbard models, we show that the interactions can destroy, but also revive, the possibility for SAP due to an interaction-induced energy-band separation.
This separation allows to generalize single-particle processes to multiple particles and requires to consider repulsively-bound pair co-tunnelling processes~\cite{Winkler:06,Zoellner:08,cotu1,cotu2,cotu3,cotu4,cotu5,cotu6}.

In the following we will first very briefly review the technique of spatial adiabatic passage (Section \ref{Sec:SAP}) and the general solution for two interacting particles in a harmonic trap (Section \ref{Sec:Exact}). We then numerically solve the problem of SAP for interacting particles (Section \ref{Sec:Exact2}) and discuss the limiting cases for weak and strong interactions using the Hubbard model (Sections \ref{Sec:BH} and \ref{Sec:FH}). The importance of level crossings and non-adiabatic evolution is discussed in Section \ref{Sec:cross} before we conclude.

\begin{figure}
\includegraphics[width=0.99 \linewidth]{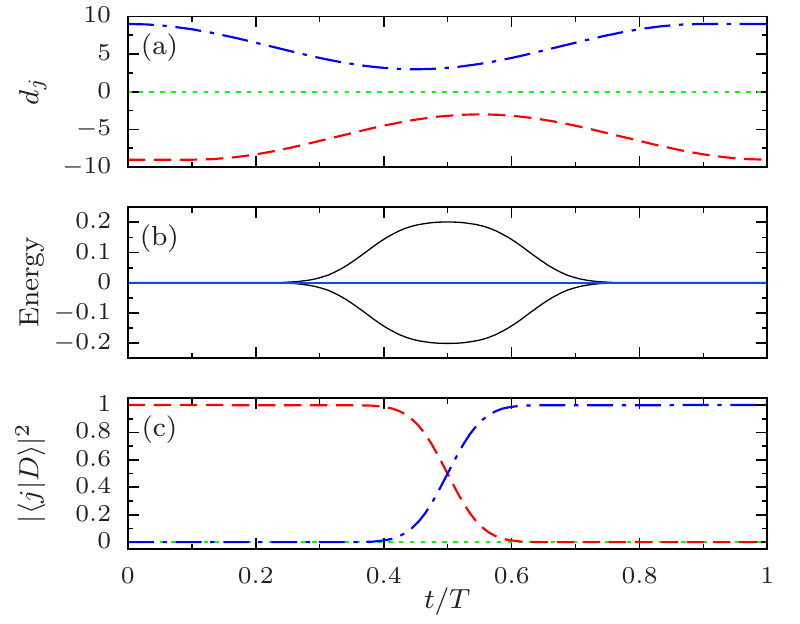}
\caption{(Color online)
(a)
Positions of the three harmonic well minima for the SAP protocol as used in our simulations
(dashed red: $d_L$, dotted green: $d_M$, dot-dashed blue: $d_R$).
The initial (and final) distance between wells is $d_{\textrm{max}}=9$, the minimum distance is $d_{\textrm{min}}=3$ and the time delay between the two approaches is $T/10$.
(b)
Energy eigenvalues of the single-particle Hamiltonian (\ref{eq:H0}), with the one corresponding to $\ket{D}$ displayed in blue (gray).
(c)
Coefficients of $\ket{D}$ in the $\{ \ket{j} \}$ basis
(dashed red: $\ket{l}$, dotted green: $\ket{m}$, dot-dashed blue: $\ket{r}$).
}
\label{fig:TLAO}
\end{figure}

\section{Single-particle SAP}
\label{Sec:SAP}

The SAP protocol for a single atom involves three degenerate localized trapping states $\ket{j}$ with $j = l$, $m$ and $r$ (for left, middle and right), which are centered at positions $d_L<d_M<d_R$.
The system is described by the Hamiltonian
\begin{align}
H_0 = \Omega\LM \ket{l}\bra{m} + \Omega\MR \ket{m}\bra{r} + \textrm{h.c.}\;, \label{eq:H0}
\end{align}
where the time-dependent nearest-neighbor couplings $\Omega_{jj'}$ are controled by the distance between the traps~\cite{eck:04}.
Note that throughout the paper we use dimensionless units where $\hbar$, atomic masses, and trapping frequencies are equal to  1.
One of the eigenstates of $H_0$ is the so-called \textit{dark state}~\cite{ber:98}, 
\begin{align}
\ket D = \cos \theta \ket l - \sin \theta \ket r,
\end{align}
with $\tan\theta=\Omega\LM/\Omega\MR$ and
SAP describes the transport of a particle from $\ket l$ to $\ket r$ following $\ket{D}$ by changing $\theta$ from $0$ ($\Omega\MR \gg \Omega\LM$) to $\pi/2$ ($\Omega\MR \ll \Omega\LM$).
This is achieved with the trap movement shown in Fig.~\ref{fig:TLAO}(a), which maintains an energy gap between the dark state and the two other eigenstates of the order of $\sqrt{\Omega\LM^2+\Omega\MR^2}$ (see Fig.~\ref{fig:TLAO}(b)) and changes the dark state from $\ket{l}$ at $t=0$ to $\ket{r}$ at final time $t=T$ (see Fig.~\ref{fig:TLAO}(c)).
To avoid excitations and ensure that SAP succeeds, the whole process needs to be carried out adiabatically.

\section{Exact model for two particles}
\label{Sec:Exact}
For simplicity, we will use a one-dimensional model of two interacting bosons in a triple-well potential, even though our results can easily be extended to higher dimensions.
The Hamiltonian is then given by
\begin{align}
H= \sum_{k=1}^2 \left( -\frac{1}{2} \frac{\partial^2}{\partial x_k^2} + V(x_k,t) \right) + g \delta(x_1-x_2), \label{eq:H} \end{align}
where $x_k$ is the position of the $k$-th atom.
The trapping potential $V$ is modeled by a piecewise harmonic triple well~\cite{eck:04} with minima located at $d_L$, $d_M,$ and $d_R$ (see Fig.~\ref{fig:TLAO}(a)).
The last term describes the contact interaction (of strength $g$) between the atoms~\cite{Busch:98}.

From here on, we refer to the ground state of two interacting bosons in each harmonic well
as $\ket{j}$ ($j=L,M,R$), whose energy $E_g$ is related to $g$ by~\cite{Busch:98}
\begin{align}
g=- \frac{2\sqrt{2} \Gamma(1-E_g/2)}{\Gamma((1-E_g)/2)},
\end{align}
where $\Gamma(x)$ is the gamma function.
Throughout the paper we will discuss the whole range of $E_g$: from the non-interacting case ($g=0$, $E_g=1$) to the Tonks--Girardeau (TG) limit ($g \rightarrow \infty$, $E_g=2$).
Our objective is to find an eigenstate of $H$, analogous to the single-particle dark state (tending to $\ket{L}$ and $\ket{R}$ at initial and final times) which allows for two-particle SAP transport between these states.

\section{SAP for interacting particles}
\label{Sec:Exact2}
\begin{figure}
\includegraphics[width=0.99 \linewidth]{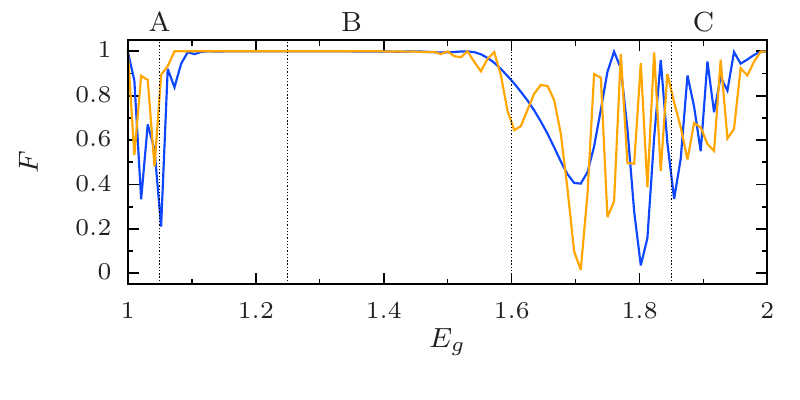} 
\caption{(Color online)
Final population in state $\ket{R}$ as a function of $E_g$ after the two-particle SAP protocol is carried out over a total time $T=4000$ (blue/dark gray) or $T=12000$ (orange/light gray).
Dotted vertical lines indicate energies for which the spectrum is shown in Figs.~\ref{fig:HB} and \ref{fig:HF}.
}
\label{fig:fid}
\end{figure}

With the system initially in state $\ket{L}$, we numerically integrate the time-dependent Schr\"odinger equation with the Hamiltonian (\ref{eq:H}) while changing the positions of the traps as shown in Fig.~\ref{fig:TLAO}(a).
We then calculate the SAP fidelity $F$ (population of $\ket{R}$ at final time $t=T$) for $E_g$ between 1 and 2.
The results for two different values of $T$ are shown in Fig.~\ref{fig:fid}, where one can identify several regions of interest.

The process succeeds for the non-interacting case ($E_g = 1$), where the particles are independent and each undergoes single-particle SAP.
Full transfer also occurs in the TG limit ($E_g = 2$), which can be easily understood since the bosonic atoms can be treated as independent fermions occupying the two lowest single-particle energy levels, and each of them can then be treated with a Hamiltonian similar to Eq.~(\ref{eq:H0})~\cite{Loiko:11}. 
The fidelity drops sharply for energies near these extreme values, regions A and C in Fig.~\ref{fig:fid}, where the population transfer is only partial and depends on $T$.
However, from $E_g\simeq 1.12$ to $E_g\simeq 1.45$ (region B) a plateau appears where $F>0.998$.

In the rest of this work, we will analyze these three regions in detail by diagonalizing the exact Hamiltonian of the system.
We will also introduce two Hubbard Hamiltonians, for weak and strong interactions,
whose formalism will allow us to study the structure of the eigenstates and give additional insight into the importance of co-tunneling processes.

\section{Weak interactions}
\label{Sec:BH}

Restricting the analysis to two-particle states in the lowest Bloch band, the system can be modeled for weak interactions using a finite-size Bose--Hubbard Hamiltonian~\cite{Jaksch:05}
\begin{align} \label{eq:BH}
H_{B} &=
\sum_{j=L,M,R} \left[ \frac U 2 n_{j} (n_{j} -1) + \epsilon_0 n_{j} \right] \\
& + \Omega\LM \left(b^\dagger_{L} b_{M}+b^\dagger_{M} b_{L}\right) 
+ \Omega\MR  \left(b^\dagger_{M} b_{R}+b^\dagger_{R} b_{M}\right) \nonumber\\
& + \Omega\co\LM  \left(b^{\dagger 2}_{L} b_{M}^{2}+b^{\dagger 2}_{M} b_{L}^{2}\right) 
+ \Omega\co\MR  \left(b^{\dagger 2}_{M} b_{R}^{2}+b^{\dagger 2}_{R} b_{M}^{2}\right) ,\nonumber
\end{align}
where $b^\dagger_j$ and $b_j$ are the creation and annihilation operators for a boson in the ground state of well $j$ and $n_j = b^\dagger_j b_j$ is the number operator.
The onsite interaction is described by $U$ and the ground state energy is $\epsilon_0=1/2$. 
Single-particle tunneling rates ($\Omega_{jj'}$) and two-particle co-tunneling rates ($\Omega\co_{jj'}$) between wells $j$ and $j'$ are numerically calculated via the Gram--Schmidt orthonormalization procedure~\cite{Loiko:11} using the single-particle~\cite{eck:04} and two-particle~\cite{Busch:98} wave functions, respectively.

When the two traps are close to each other, $\Omega_{jj'}$ and $\Omega\co_{jj'}$ are of the same order of magnitude, but for weak interactions single-particle tunneling dominates due to its larger amplitude.
For stronger interactions, however, single-particle tunneling becomes off-resonant, unlike co-tunneling which dominates because it remains resonant.
While co-tunneling is not usually considered, it is crucial to understand the behavior observed in the previous section.
A study of $H_B$ (with real-time simulations, eigenvalues and eigenstates) with and without co-tunneling terms can be found in the appendix.

\begin{figure}
\includegraphics[width=0.99 \linewidth]{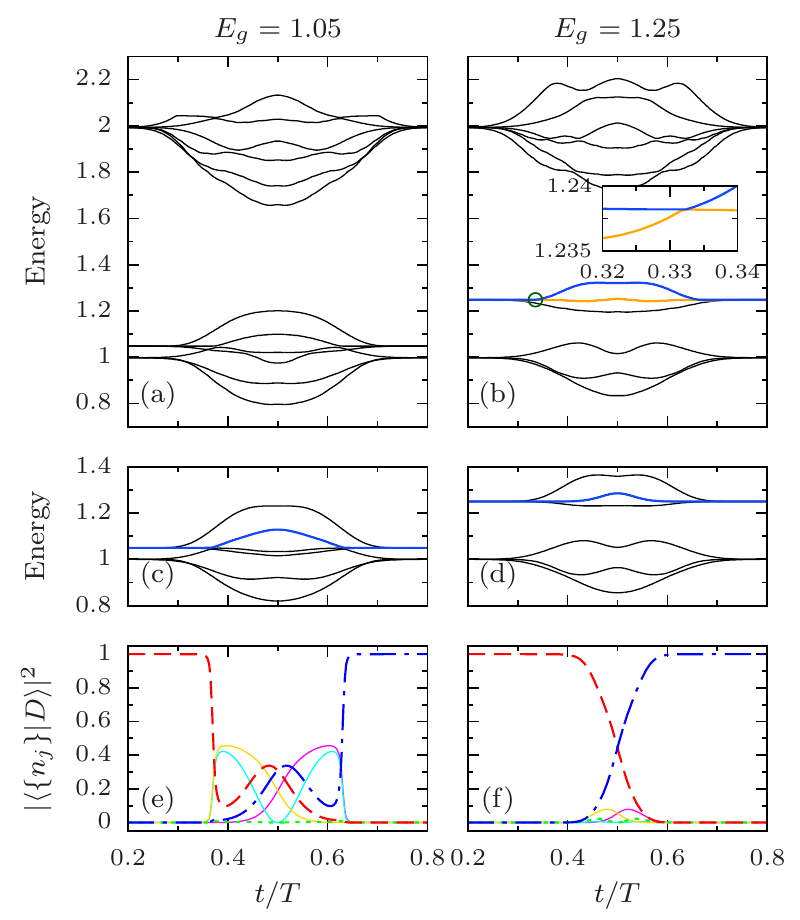} 
\caption{(Color online) (a,b)
Lowest 12 eigenvalues of $H$ for the two-particle SAP scheme with the trap moving sequence of Fig.~\ref{fig:TLAO}(a) for (a) $E_g=1.05$ and (b) $E_g=1.25$.
In (b) the energy of the dark state (asymptotically $\ket{L}$ and $\ket{R}$) and the state with which it couples the most are marked in blue (dark gray) and orange (light gray), respectively.
The inset shows a zoom-in of the marked crossing between these two states (marked with a circle) which is analyzed in Sec.~\ref{Sec:cross}.
(c,d) Eigenvalues of $H_B$ for the same parameters as (a,b), with the energy of the dark state drawn in blue (gray).
(e,f) Coefficients in the Fock basis $\ket{\{n_{j}\}}$ of the dark states in (c,d) 
(dashed red corresponds to $\ket{L}$,
dotted green to $\ket{M}$,
dot-dashed blue to $\ket{R}$,
and the solid lines to states where the two atoms are in different traps).
\label{fig:HB}}
\end{figure}

In order to understand regions A and B in Fig.~\ref{fig:fid} we diagonalize numerically both $H$ and $H_B$ at any point during the SAP evolution.
The spectra of $H$ for $E_g = 1.05$ and 1.25 are shown in Fig.~\ref{fig:HB}(a,b), and consist of three distinct bands.
The lowest band contains three states (where the two atoms sit in different wells), and for large trap separations (i.e., initial and final times) has energy $1$.
Not discussed here, this band contains an eigenstate which allows the adiabatic transfer of an atomic hole between the outermost wells for strong enough interactions~\cite{Benseny:10}.
The middle band (also with three states) is the one of interest to us because it only involves states with the two atoms in the same trap, $\ket{j}$, with energies around $1+U = E_g$.
The higher band, with energies around 2, consists of six states which correspond to states where the atoms sit in the ground state and first excited states of different traps.
For the same parameters, the spectrum of $H_B$ is shown in Fig.~\ref{fig:HB}(c,d).
These spectra show two bands, corresponding (with good agreement) to the two lower bands of the exact Hamiltonian we have just discussed.
Higher bands do not appear because only the lowest Bloch band has been considered in $H_B$.

The band around $E_g$ has a similar structure to the single-particle SAP spectrum (cf. Fig.~\ref{fig:TLAO}(b)) and it contains an eigenstate, drawn in blue (dark gray) in Fig.~\ref{fig:HB}(c,d), which allows for SAP transport because of its dark-state-like structure, shown in Fig.~\ref{fig:HB}(e,f).
It is important to remark that this state is only present if co-tunneling is considered in $H_B$, see the appendix.
The SAP transport fails for weak interactions, Fig.~\ref{fig:HB}(a,c,e), because when tunneling becomes relevant the two bands overlap and dynamically following the dark state is hard due to the presence of level crossings.
We can then see that the appearance of the plateau in Fig.~\ref{fig:fid} for stronger interactions, i.e., $U\gtrsim\sqrt{2}\Omega\sim0.15$, is due to the two bands remaining separated during the whole process, see Fig.~\ref{fig:HB}(b,d).
This allows for the SAP process to be successful again because the band with the dark state resembles the single-particle SAP spectrum, cf. Fig.~\ref{fig:TLAO}(b).
The dark state has a similar shape to the single-particle one but for the states $\ket{j}$ coupled through co-tunneling (compare Figs.~\ref{fig:HB}(f) and \ref{fig:TLAO}(c)).

\section{Strong interactions}
\label{Sec:FH}

One expects then that for stronger interactions the process will keep working, until $E_g$ approaches the band with energies around 2, which will cause the fidelity to drop again in region C of Fig.~\ref{fig:fid}.
Because of the fermionic behavior of the TG gas, this regime is best modeled by restricting the system to two-particle Fock states with one atom in each of the two lowest Bloch bands and using a finite-size Fermi--Hubbard Hamiltonian~\cite{Jaksch:05}
\begin{align}
H_{F}=
&\sum_{j=L,M,R} \left[ U n_{j0} n_{j1} + \sum_{i=0,1} \epsilon_i n_{ji} \right] \\ 
& + \sum_{i=0,1} \left[ \Omega^{(i)}\LM a^\dagger_{Li} a_{Mi} + \Omega^{(i)}\MR a^\dagger_{Mi} a_{Ri} + \textrm{h.c.} \right] \nonumber \\
& + \Omega\co\LM a^{\dagger}_{L0} a^{\dagger}_{L1} a_{M0} a_{M1} + \textrm{h.c.} \nonumber \\
& + \Omega\co\MR a^{\dagger}_{M0} a^{\dagger}_{M1} a_{R0} a_{R1} + \textrm{h.c.} \nonumber
\end{align}
Here $a^\dagger_{ji}$ and $a_{ji}$ are the fermionic creation and annihilation operators for a particle at energy level $i$ of well $j$, $\Omega^{(i)}_{jj'}$ are the tunneling frequencies at level $i$, $n_{ji} = a^\dagger_{ji} a_{ji}$, and $\epsilon_i=1/2+i$.
The onsite interaction is given by $U$, which is now negative since two bosons interacting repulsively with finite strength have less energy than two non-interacting fermions.
In other words, the ground state energy is now $\epsilon_0+\epsilon_1+U=2-|U| = E_g$.

\begin{figure}
\includegraphics[width=0.99 \linewidth]{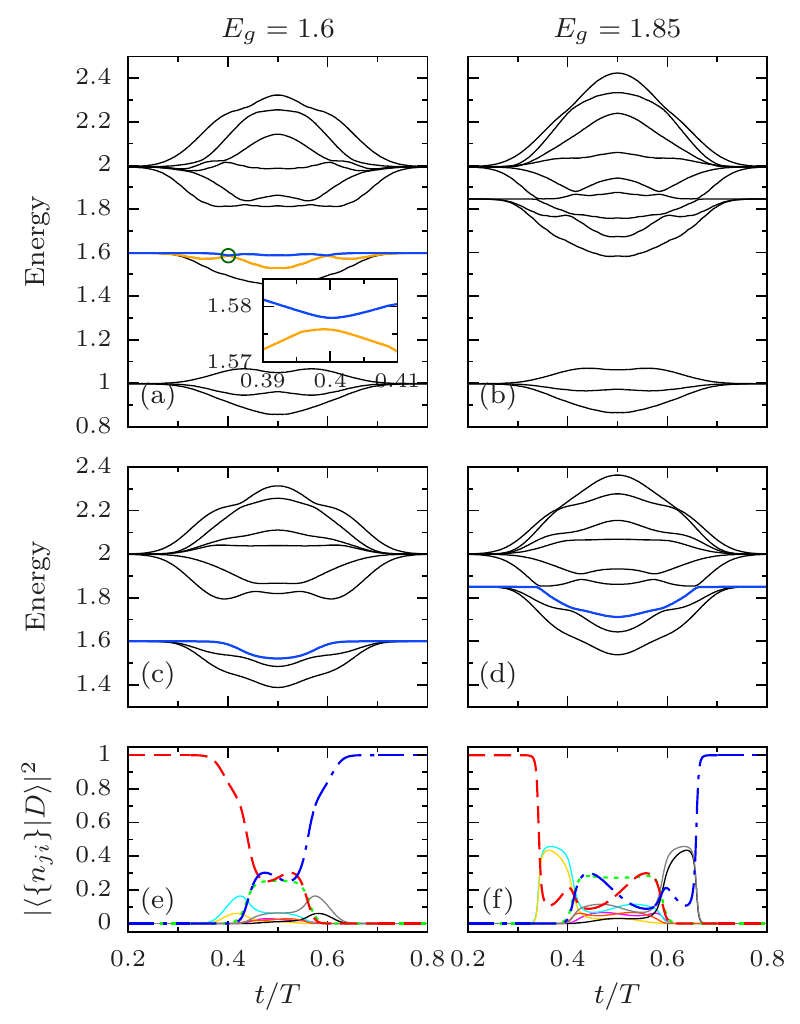} 
\caption{(Color online) (a,b)
Lowest 12 eigenvalues of $H$ for the two-particle SAP scheme with the trap moving sequence of Fig.~\ref{fig:TLAO}(a) for (a) $E_g=1.6$ and (b) $E_g=1.85$.
In (a) the energy of the dark state (asymptotically $\ket{L}$ and $\ket{R}$) and the state with which it couples the most are marked in blue (dark gray) and orange (light gray), respectively.
The inset shows a zoom-in of the marked crossing between these two states (marked with a circle) which is analyzed in Sec.~\ref{Sec:cross}.
(c,d) Eigenvalues of $H_F$ for the same parameters as (a,b), with the energy of the dark state drawn in blue (gray).
(e,f) Coefficients of the dark state in (c,d) the Fock basis $\ket{\{n_{ji}\}}$ (color coding is the same as in Figs.~\ref{fig:HB}(e,f)).
}
\label{fig:HF}
\end{figure}

We show in Fig.~\ref{fig:HF}(a,b) the spectrum of $H$ for $E_g=1.6$ and 1.85, and one can see the same three-band structure as in Fig.~\ref{fig:HB}.
For the same values of $E_g$, the spectrum of $H_F$, Fig.~\ref{fig:HF}(c,d), consists of two bands: around $2-|U|=E_g$ and 2, in good agreement with the exact results.
Once again, one state in the band around $E_g$ (in blue/gray) has the form of a dark state, see Fig.~\ref{fig:HF}(e,f).
As $E_g$ increases, the two bands start to overlap and crossings appear, leading to the dark state involving states of the upper band, see Fig.~\ref{fig:HF}(b,d), which perturb the adiabatic transfer as seen in region C.
It is worth noting that $H_F$ contains a dark state even without the co-tunneling term.

\section{(A)diabatic requirements}
\label{Sec:cross}

Thus far we have seen that our system's Hamiltonian contains a band with energies around $E_g$ which consists of a three-level system of states $\ket{j}$, coupled by repulsively-bound pair tunneling.
Thanks to the atomic interaction, this band can be isolated from the rest of two-particle states, and allows the transport to succeed.
From the size of the two bands of $H_F$, one would expect that process should not be affected by level crossings until around $E_g\sim1.7$.
However, the plateau in Fig.~\ref{fig:fid} only extends until $E_g\sim1.5$.
In order to understand this we will look back on the spectrum of the exact Hamiltonian spectra in Figs.~\ref{fig:HB}(b) and \ref{fig:HF}(a).
For both spectra we have highlighted in blue (dark gray) the energy of the dark state, i.e., the state which at initial and final times is $\ket L$ and $\ket R$, respectively, and therefore the state we are initially following. 
In contrast to the Hubbard Hamiltonians, this dark state crosses another eigenstate (shown in orange/light gray) twice, creating a finite probability for the system to leave the dark state. For $1.1 \lesssim E_g \lesssim 1.8$ these are the only relevant crossings affecting the dark state, and we examine them carefully in the following.
However, as the crossings are actually avoided crossings, it is clear that the speed at which they are passed will determine the adiabaticity and the success of the transfer~\cite{lan81,harkonen06}.

\begin{figure}
\includegraphics[width=0.99 \linewidth]{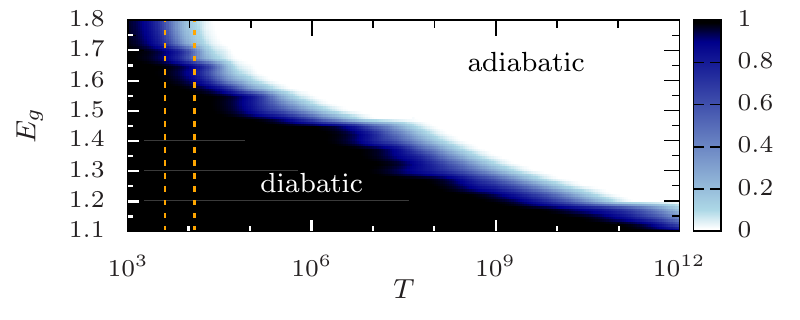}
\caption{(Color online)
Probability of transition $p_{i\rightarrow j}$ at the crossing between the eigenstates shown in the insets of Figs.~\ref{fig:HB} and \ref{fig:HF} for different total times $T$ and energies $E_g$.
Dashed vertical lines indicate the total times used in Fig.~\ref{fig:fid}.
}
\label{fig:trans}
\end{figure}

The probability for a system following an eigenvector $\ket{i(t)}$ to be excited into another one $\ket{j(t)}$ between times $t_0$ and $t_f$ can be approximated by~\cite{mes61}
\begin{align}
\label{eq:trans}
p_{i\rightarrow j} \simeq \frac{ \left| \int_{t_0}^{t_f} \bra{j(t)} \frac{d}{dt} \ket{i(t)} e^{i \int_{t_0}^{t} \left( E_j(\tau)-E_i(\tau)\right) d\tau} dt \right|^2}{ \left| \int_{t_0}^{t_f} \bra{j(t)} \frac{d}{dt} \ket{i(t)} dt \right|^2},
\end{align}
where $E_k(t)$ is the energy of state $\ket{k(t)}$.
The denominator was added from the original expression for normalization purposes and represents the probability for the state to be excited for an infinitely fast process.
Full transfer through the SAP protocol can be achieved if both crossings are passed either adiabatically ($p_{i\rightarrow j} = 0$, following always the dark state) or completely diabatically ($p_{i\rightarrow j} = 1$, following the orange (light gray) state between the crossings).
For intermediate values of $p_{i\rightarrow j}$, the system's actual state will be distributed between different eigenstates and the transfer will not be complete.

In Fig.~\ref{fig:trans} we show the calculated transition probabilities for the gap as a function of $T$ and $E_g$.
One can see that for the timescales of $T$ used in Fig.~\ref{fig:fid}, the transfer is completely diabatic for $E_g=1.25$.
For $E_g=1.6$, however, $p_{i\rightarrow j}$ starts to falls to $0.99$ ($0.96$) for $T=4000$ ($12000$), and even lower for higher $E_g$.
Therefore, the plateau ends because of the increased size of the gap in the avoided crossing, which no longer allows for diabatic passage.
Moreover, it can be seen that the length and position of the plateau is $T$-dependent, since the transition from a diabatic to an adiabatic process happens at very different time scales, which depend on $E_g$.
For longer time scales, e.g.~of the order of $10^7$, two plateaus would appear in region B: one from $E_g \sim 1.1$--$1.4$ in which the transfer in this crossing would be completely diabatic, and one at $E_g \sim 1.5$--$1.7$ where the transfer would be adiabatic.

\section{Conclusion}

In this work we have studied the spatial adiabatic passage protocol for a system of interacting bosons over the entire range of repulsive interactions.
We have found that, in addition to the trivial cases for non-interacting and infinitely strongly interacting particles, a large and continuous region for intermediate interactions exist over which high fidelities can be obtained.
This is due to the fact that for intermediate values of $E_g$ a decoupled energy band appears, which possess a dark state facilitated by two-particle co-tunneling.
However, when this band overlaps with other energy bands, the appearance of level crossings prevents the robust use of the dark state.
This behavior is generic to any multi-well setting and not specific to SAP.
It is worth noting that the above effect is limited to systems where no phonon modes exist and therefore does, for example, not apply to SAP in quantum dot systems.

\begin{acknowledgments}

This work was supported by the Okinawa Institute of Science and Technology Graduate University.
We thank Irina Reshodko for helpful discussions.

\end{acknowledgments}

\appendix

\section*{Appendix: Justification of the co-tunneling terms}

\begin{figure}
\includegraphics[width=0.99 \linewidth]{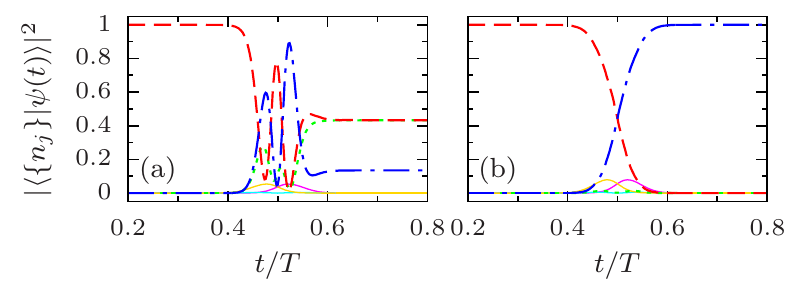} 
\caption{(Color online)
Populations of the eigenstates from integrating the Schr\"odinger equation with the Hamiltonian $H_B$ for $E_g=1.25$ (a) without and (b) with the co-tunneling terms.
The initial state is $\ket{\psi(t=0)}=\ket{L}$ and the tunneling rates are calculated numerically with the time-dependent trap positions in Fig.~\ref{fig:TLAO}(a).
Color coding is the same as in Figs.~\ref{fig:HB}(e,f).
}
\label{fig:HB_rt}
\end{figure}

\begin{figure}
\includegraphics[width=0.99 \linewidth]{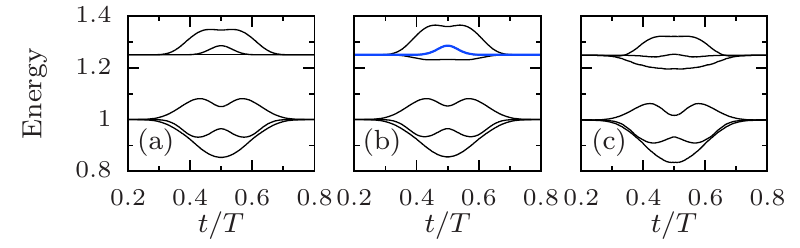}
\caption{(Color online)
Spectra of (a) the exact Hamiltonian (lowest six eigenstates), and $H_B$ without (b)  and with (c) the co-tunneling terms for the SAP process as a function of time for $E_g = 1.25$ ($U = 0.25$).
The energy of the dark state in (b) is shown in blue (gray).
}
\label{fig:compare_spectra}
\end{figure}

\begin{figure}
\includegraphics[width=0.99 \linewidth]{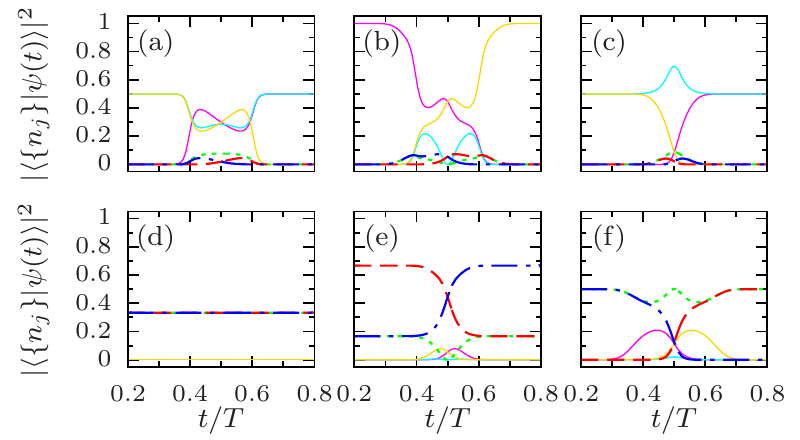}
\caption{(Color online)
Populations of the eigenstates of the $H_B$ without the co-tunneling terms in the Fock basis for the SAP process as a function of time.
States are ordered with increasing energy as (a) to (f), and use the same color coding as Figs.~\ref{fig:HB}(e,f).
}
\label{fig:diag_without}
\end{figure}

\begin{figure}
\includegraphics[width=0.99 \linewidth]{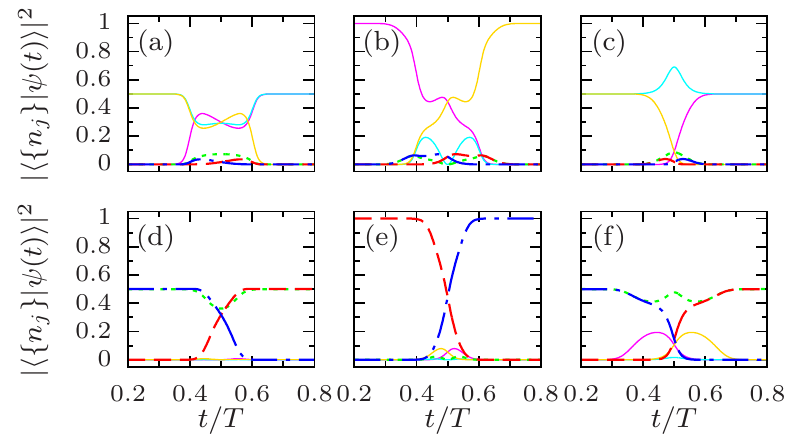}
\caption{(Color online)
Same as Fig.~\ref{fig:diag_without} but including the co-tunneling terms in $H_B$.
}
\label{fig:diag_with}
\end{figure}

In this appendix we will examine the validity of the Bose--Hubbard Hamiltonian in greater detail and confirm the importance of the co-tunneling terms. For this we have performed numerical simulations of the time-dependent Schr\"odinger equation using $H_B$ for $E_g=1.25$ ($U=0.25$) and,
because this falls inside region B of Fig.~\ref{fig:fid}, the process is expected to work.
The results, shown in Fig.~\ref{fig:HB_rt}, compare the cases where the co-tunneling terms in Eq.~\eqref{eq:BH} are either neglected or considered and one can immediately notice that the simulation that does not consider the co-tunneling terms gives incomplete transfer to state $\ket{R}$.
However, when one considers the co-tunneling terms, the transfer is complete, which establishes that these terms are crucial in order to explain the two-particle dynamics.

To shed more light on this we have also computed the spectrum of $H_B$, without and with co-tunneling, and that of the exact Hamiltonian during SAP.
The results are shown in Fig.~\ref{fig:compare_spectra}, and one can see that the inclusion of the co-tunneling terms leads to $H_B$ approximating the exact spectrum more closely.
Furthermore, we have computed the populations of the eigenstates in the Fock basis without (Fig.~\ref{fig:diag_without}) and with (Fig.~\ref{fig:diag_with}) co-tunneling terms.
States in the lower band, depicted in (a-c), are mostly composed by states where the atoms are in separate traps.
Because these state do not couple through co-tunneling, both their structure and energies coincide in Figs.~\ref{fig:diag_without} and \ref{fig:diag_with}.
It is noteworthy that the state in (b) corresponds to another kind of dark state that transfers an atomic hole from the right well to the left well~\cite{ben12}.
The states in the higher band, shown in (d-f), clearly differ for the two Hamiltonians.
When the co-tunneling terms are absent, no dark state that allows transitions from $\ket{L}$ to $\ket{R}$ exists in $H_B$, see Fig.~\ref{fig:diag_without}. However, it is clearly present when taking into account co-tunneling,  see Fig.~\ref{fig:diag_with}(e) (the energy of the dark state is shown in blue (gray) in Fig.~\ref{fig:compare_spectra}(b)).

Therefore we have clearly established that a dark state that allows transfer between $\ket{L}$ and $\ket{R}$ is only present in $H_B$ if the co-tunneling terms are considered.

\end{document}